\def\Ryear{2013} 
\def\Rno{11}      
\documentclass{petraport}
\usepackage{amsfonts,algorithm,algorithmic,graphicx,hyperref,amsmath}
\newcommand{\unit}[1]{\ensuremath{\, \mathrm{#1}}}

\begin{document}
\RaportTitle{Simulations of gamma quanta scattering in a single module \\ of the J-PET detector}
\RaportAuthors{K. Szymański\textsuperscript{1}, 
P.~Moskal\textsuperscript{1}, 
T.~Bednarski\textsuperscript{1}, 
P.~Białas\textsuperscript{1}, 
E.~Czerwiński\textsuperscript{1}, 
K.~Giergiel\textsuperscript{1}, 
Ł.~Kapłon\textsuperscript{1,2}, 
A.~Kochanowski\textsuperscript{2}, 
G.~Korcyl\textsuperscript{1}, 
J.~Kowal\textsuperscript{1}, 
P.~Kowalski\textsuperscript{3}, 
T.~Kozik\textsuperscript{1}, 
W.~Krzemień\textsuperscript{1}, 
M.~Molenda\textsuperscript{2}, 
I.~Moskal\textsuperscript{1},
Sz.~Niedźwiecki\textsuperscript{1}, 
M.~Pałka\textsuperscript{1}, 
M.~Pawlik\textsuperscript{1}, 
L.~Raczyński\textsuperscript{3}, 
Z.~Rudy\textsuperscript{1}, 
P.~Salabura\textsuperscript{1}, 
N.G.~Sharma\textsuperscript{1}, 
M.~Silarski\textsuperscript{1}, 
A.~Słomski\textsuperscript{1}, 
J.~Smyrski\textsuperscript{1}, 
A.~Strzelecki\textsuperscript{1}, 
P.~Witkowski\textsuperscript{1}, 
W.~Wiślicki\textsuperscript{3}, 
M.~Zieliński\textsuperscript{1}, 
N.~Zoń\textsuperscript{1}
}
\RaportInstitute{\textsuperscript{1}Institute of Physics, Jagiellonian University, 30-059 Cracow, Poland\\
\textsuperscript{2}Faculty of Chemistry, Jagiellonian University, 30-060 Cracow, Poland\\
\textsuperscript{3}Świerk Computing Centre, National Centre for Nuclear Research, 05-400 Otwock-Świerk, Poland}

\section{Abstract}
This article describes simulations of scattering of annihilation gamma quanta 
in a strip of plastic scintillator. Such strips constitute
basic detection modules in a newly proposed Positron Emission Tomography 
which utilizes 
plastic scintillators instead of inorganic crystals.
An algorithm simulating chain of Compton scatterings was elaborated and series of simulations have been conducted
for the scintillator strip with the cross section of  5~mm x 19~mm.
Obtained results indicate that secondary interactions occur only in the case of about 8$\%$ of events
and out of them only 25$\%$ take place in the distance larger than 0.5~cm from the primary interaction.  
It was also established 
that light signals produced at primary and secondary interactions overlap with the delay
which distribution is characterized by FWHM of about 40~ps.

\section{Introduction}
Recently a novel solution for the Positron Emission Tomography scanner
was proposed which utilizes plastic scintillators as detectors
for the annihilation quanta~\cite{BAMS,BAMS1,BAMS2}. 
A single detection unit of this detector
is built out of a scintillator strip read out on both sides by photomultipliers. 
The position of the interaction of the gamma quanta inside
a strip is determined based on the shape and time of photomultiplier signals. 
The shape of signals may be distorted by the secondary 
scattering of gamma quanta inside the scintillator. This is because a secondary scattering 
creates an additional light signal and as a result a light pulse reaching the photomultiplier 
is composed of overlaping signals originating 
from points of primary and secondary interactions.

In this article we estimate influence of the secondary scattering on the quality of hit position reconstruction
in the PET detectors based on the plastic scintillators.
To this end a dedicated simulation programme was elaborated and series of simulations have been performed. These allowed to determine multiplicity distributions of annihilation quanta interaction inside a scintillator strip, as well as distributions of distance between points of scattering.
The obtained results are then interpreted 
in view of the distortion of spatial and temporal resolution of the detector due to the secondary interactions 
of registered gamma quanta. The main simulation algorithm is elaborated assuming that in plastic scintillators
the annihilation gamma quanta with energy of about 511~keV, undergo a Compton scattering only.

\section{Algorithm}
Main aim of the simulation was an estimation of number of gamma quanta interactions in a scintillator strip 
and determinantion of spatial distribution of scattering centers within the scintillator volume. 
For this purpose, we assume that gamma quantum originates in a certrain initial position
with a user--defined four--momentum vector. Description of scintillator is also provided by user and consists of: 
attenuation constant at energy for \(511 \unit{keV}\) and dimensions of scintillator cuboid. 
From this input data attenuation constant at any energy is extrapolated using following formula:
\begin{equation}
\lambda(E)=\frac{\sigma(E)}{\sigma(E_0)} \lambda(E_0)
\end{equation}
Where \(\lambda\) denotes inverse of attenutation length, and 
\(\sigma\) stands for total cross section of gamma quantum at given energy \(E\).
The values of cross sections have been extracted from~\cite{nistxcom}.
A given number of events, \(N\), and maximum number of interactions in one event, \(k\), is also passed to algorithm, 
which may be in general described by a following scheme:
\begin{algorithmic}
 \WHILE{number of events \( \le \) N}
    \STATE position \(\gets\) initial position
    \STATE direction \(\gets\) initial direction 
    \STATE energy \(\gets\) initial energy
    \WHILE{number of interactions \( \le \) k}
   
     \STATE \(\lambda \gets \) \(n_e\) \(\cdot\) total cross section
      \STATE length \(\gets\) random length of PDF \(\rho(x>0)=\lambda \exp(-x \lambda)\)
      \STATE position \(\gets\) position + direction \(\cdot\) length
      \IF{position not in scintillator}
       \STATE end event
      \ENDIF 
      \STATE polar angle \(\gets\) random angle from Klein-Nishina PDF
      \STATE azimuthal angle \(\gets\) random angle from izotropic PDF
      \STATE \emph{scatter} direction vector using generated angles
      \STATE print position, deposited energy, time
    \ENDWHILE
  \ENDWHILE
  \end{algorithmic}
As a result, collection of events containing ordered set of interactions, is obtained. 

\subsection{Generating random data from desired probability density function}
The generation of varaiables used in the programme according to the given Probability Density Function (PDF) 
is performed based on the 
distribution of a corresponding cumulative function, which is homogenous by definition.   
Cumulative function (\(D(\theta)\)) of the angular Probability Density Function is calculated 
based on the Klein-Nishina formula \cite{zlacalka}. 
In following equations it is factorized into \(\theta\)--independent (constant \(C\)) and \(\theta\)--dependent functions, 
and other \(\theta\)--independent function is added 
(only relative changes to \(D(0)\) are important; additive function is chosen to ensure that \(D(0)~=~0\), which simplifies calculations):
\begin{align*}
D(\theta)&=C \cdot [(-5\gamma^2+2(2\gamma+1)\gamma\cos\theta-6\gamma-2) P^2(\theta,\gamma)\\
&+2(\gamma^2-2\gamma-2)\ln(\frac 1{P(\theta,\gamma)})-2\gamma\cos\theta]\\
P(\theta,\gamma)&=\frac {1}{1+\gamma(1-\cos\theta)}\\
\gamma&=\frac{E}{m_e c^2},
\end{align*}
where \(E\) denotes energy of gamma quantum, and  \(m_e\) stands for the mass of electron.\\
This method is used only for generation of scattering angle \(\theta\). In the case of the distance \(l\) between subsequent scattering a cumulative distribution function can be derived analyticaly since PDF posesses an exponential form:
\begin{equation}
P(l)=\frac 1 \lambda \exp(-\lambda l)
\end{equation}
where \(l\ge0\). Hence, cumulative function reads:
\begin{equation}
D(l)=1-\exp(-\lambda l).
\end{equation}
Therefore, for a given random number of uniform distribution \(r\in[0,1]\):
\begin{equation}
l=-\frac{\ln (1-r)}{\lambda}.
\end{equation}

\subsection{A method of simulation of gamma quanta scattering}
\begin{figure}[!h]
\centering
\includegraphics[width=0.5\textwidth]{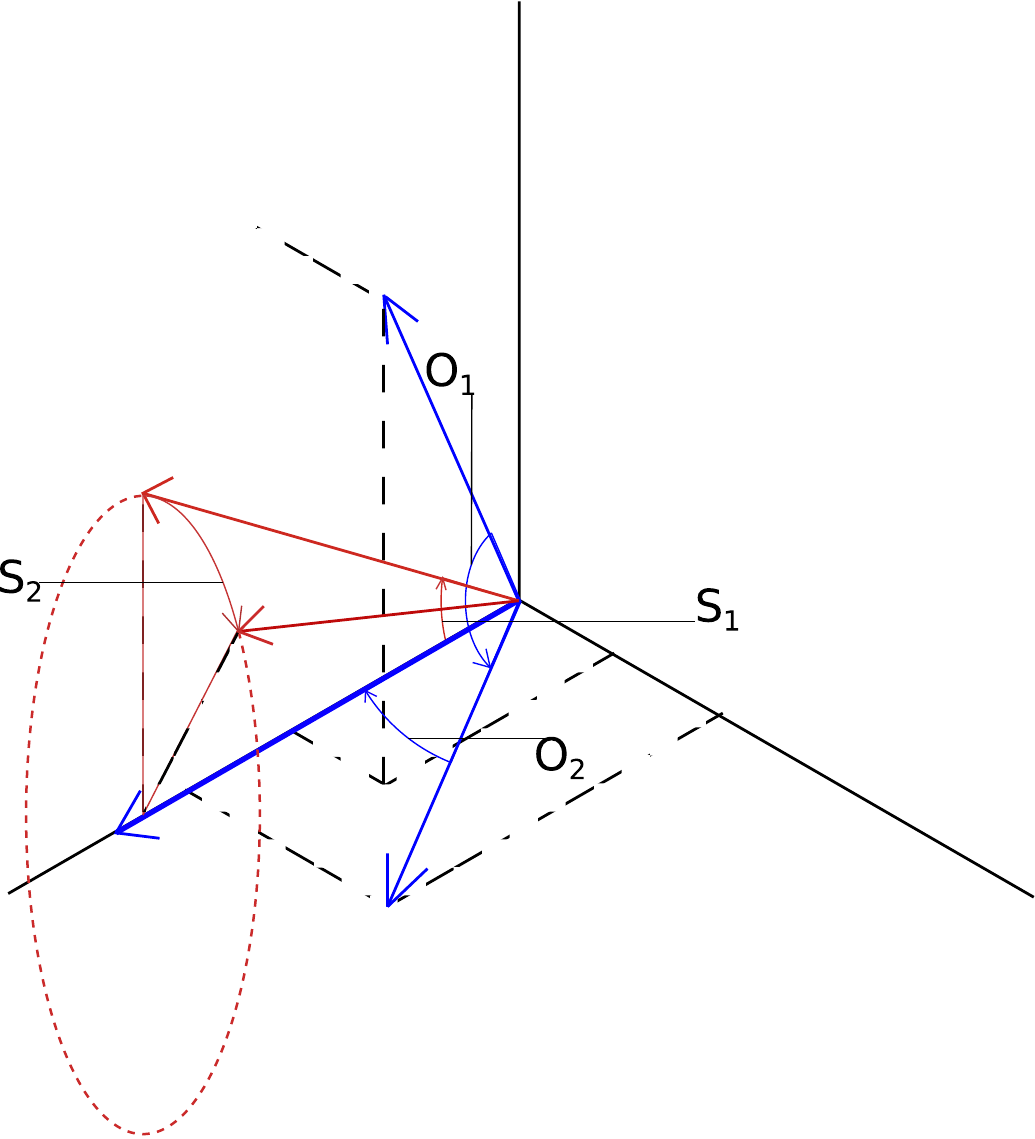}
\caption{Pictorial presentation of vector rotations applied for simulations of the Compton scattering.}
\label{scatter}
\end{figure}

In order to effectively simulate a series of scatterings, we have elaborated an algorithm in which calculations reduces to the 
rotations of vectors.
For given vector \(v\in\mathbb{R}^3\) the rotation matrix \(O_1\)~\footnote{\(O\)-matrices are changing reference frame, while \(S\)-ones describe transformation of vector in particular frame.} fullfiling the condition
\begin{equation*}
O_1 \cdot v=\begin{bmatrix}0\\v'_2\\v'_3\end{bmatrix}
\end{equation*}
is generated. Let \(O_1 \cdot v=v'\). Then, another rotation matrix \(O_2\), such that
\begin{equation*}
v''=O_2 \cdot v'=\begin{bmatrix} 0\\0\\v''_3\end{bmatrix}
\end{equation*}
is computed. Now scattering can be  simulated by rotation of vector  \(v''\) by means of matrix \(S_1\) in such way that \(S_1 \cdot v''\) forms a desired angle with z-axis. To preserve cylindrical symmetry, this vector must be rotated by rotation matrix around the z-axis \(S_2\) by a random angle. Generated vector must be re-transformed to original reference frame, and the final scattered vector is given by:
\begin{equation*}
v_{\mathrm{scatt}}=O_1^{-1} O_2^{-1} S_2 S_1 O_2 O_1 v.
\end{equation*}
The discussed rotations are visualized in Fig.~\ref{scatter}.

\subsection{Validation of algorithm assumptions}
\begin{figure}[!h]
\centering
\includegraphics[width=0.6\textwidth]{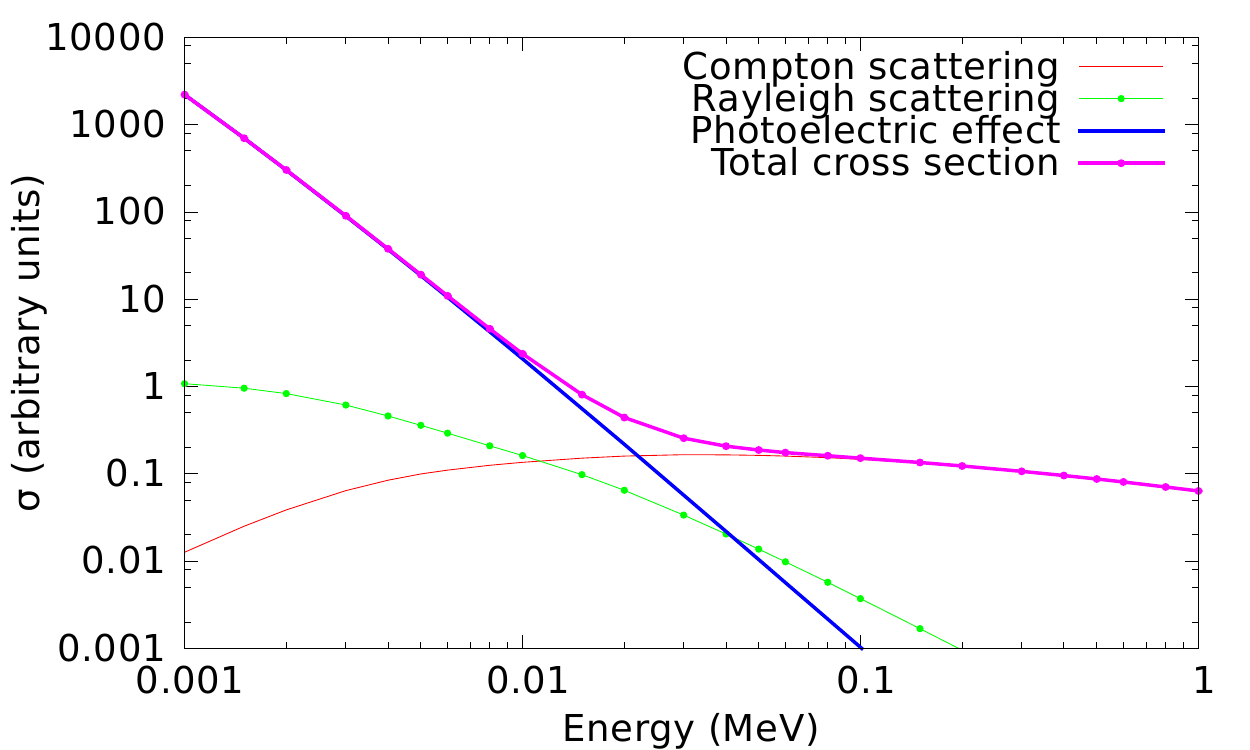}
\caption{Total cross sections of various effects for gamma quanta scattering in carbon as a function of energy~\cite{nistxcom}.}
\label{fig:tcss}
\end{figure}

The gamma quantum passing through the matter may undergo various processes among which Compton, Rayleigh and photelectric effects
are dominant for the energy range below 1 MeV. 
For simulations of the interaction of annihilation quanta in a plastic scintillator we assumed the 
dominance of the Compton effect and have neglected other processes since they can contribute significantly only for energies below 
50 keV (see Fig.~\ref{fig:tcss}). 

The energy of a gamma quantum may be expressed 
in terms of the \emph{reduced gamma quantum energy} 
\(\gamma=\frac{h \nu}{m_e c^2}\). 
Such notation is especially convenient for calculations involving annihilations quanta for which 
\(\gamma=1\). For further consideration it is useful to mention that \(\gamma\approx0.1\) for gamma quantum with energy of 50 keV. 

In the scintillator strip annihilation gamma quantum may in principle undergo many Compton scatterings and its minimum energy after  
\(n\)--th interaction may be calculated based on the iterative formula:
\begin{equation}
\gamma_{n+1}=\frac {\gamma_n}{1+2 \gamma_n}
\end{equation}

First terms of this series are \((1,0.33,0.19,0.14,0.11,0.09,\dots)\). 
It shows, that up to 4th interaction energy of gamma quantum is above \(50\unit{keV}\). 
It will be shown in the following sections that in the case of the scintillator strips of the J-PET detector
a probability that the gamma quantum scatters more than three times 
is negligible. Therefore, all efects
other than Compton scattering can be safely neglected.\\
Another assumption applied is the convexity of scintillator's shape: 
in such case if gamma quantum leaves scintillator, it will never come 
back and computations can stop. It is also certain (from definition) 
that line segment bounded by two points of interactions is inside 
scintillator's volume, which simplifies calculations of next interaction point.

\section{Results}
\begin{figure}[!h]
\centering
\includegraphics[width=0.5\textwidth]{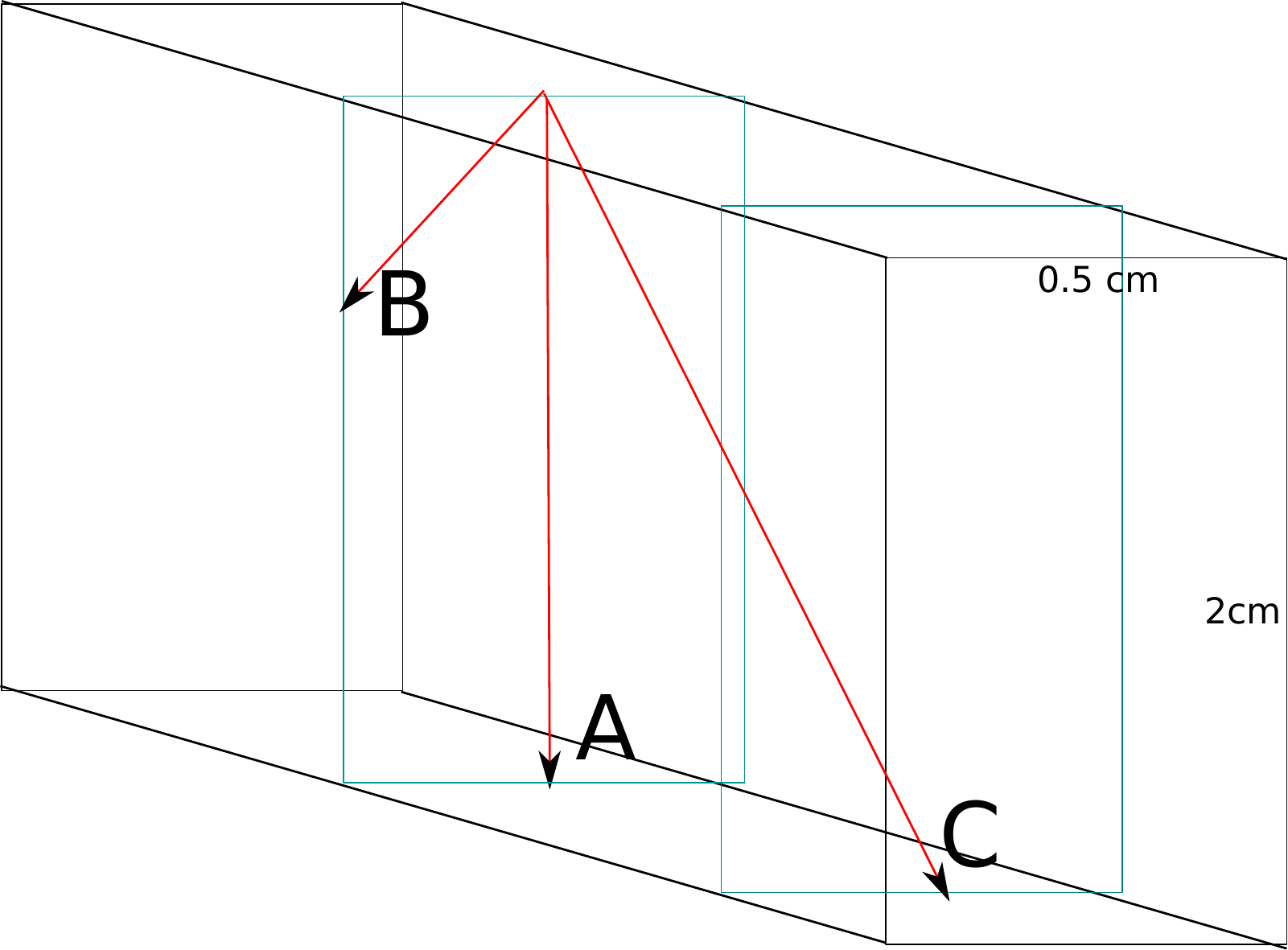}
\caption{Graphical presentation of simulation conditions. 
Red lines show initial direction of flight of gamma quanta for case A, B and C as it is indicated in the Figure.
\label{cond}}
\end{figure}

We have performed simulations assuming that the scintillator strip is a cuboid with dimensions of  
\(2000\unit{cm} \times 0.5\unit{cm} \times 2 \unit{cm}\). The cross section of the strip corresponds to the size
of the modules used for the first J-PET prototype, and its length was chosen arbitrarily large.
Starting point of gamma quanta was placed on the middle of \(2000\unit{cm} \times 0.5\unit{cm}\) wall. 
Its initial energy was set to \(m_e c^2\approx 511\unit{keV}\)~(\(\gamma=1\)). 
Number \(k\) was set to 1000 to make sure that all possible interactions were taken into account. 
Simulations were performed for three directions of the gamma quanta shown in Fig.~\ref{cond}.
Three cases denoted in Fig.~\ref{cond} as A, B, C 
have initial direction vectors \((0,0,-1)\), \((0,-1,-1)\) and \((-1,0,-1)\), respectively. 
For each direction \(10^6\) events were simulated. 
Attenuation constant 
\footnote{Derived from information in \cite{dlugosc} for BC--408, BC--420 anc BC--422 scintillators and \cite{nistxcom}.}
at \(\gamma=1\) was set to \(0.1 \unit{cm}^{-1}\).\\
\subsection{Multiplicity of interactions}
Distribution of multiplicity of interactions determined for three studied directions A, B and C is shown in Table \ref{tab:freq}.
\begin{table}[H]
\centering
\begin{tabular}{|c|c|c|c|}
\hline Level & Frequency (A) & (B) & (C)\\
\hline\hline
1 & 100 \% & 100 \% & 100 \% \\
2 & 7.9  \% & 5.7 \%& 8.3\% \\
3 & 0.66  \% & 0.44\% & 0.71 \%\\
4 & 0.06\ \% & 0.04 \% & 0.07\% \\\hline
\end{tabular}
\caption{Frequency of occurence of interactions. 
`Frequency' of level \(k\) is indicating fraction of events with \(k\) or more number of interactions.
The values are obtained taking into account only these events for which gamma quantum interacted in the scintillator.}
\label{tab:freq}
\end{table}
The result presented in this table indicates that only in about 8$\%$ of events 
the gamma quantum scatters more than once inside the scintillator strip.

\subsection{Spatial distribution of interactions points}
The points of primary interaction populates a line along the direction of the flight of the gamma quantum
with density decreasing exponentially with a distance from a point of emission. This decrease is governed by the attanuation 
length which for the plastic scintillator and annihilation gamma quanta amounts to about 10~cm~\cite{dlugosc}. 
The distribution of distances between consecutive interaction points is more complicated to describe because
the energy and hance the attanuation length of scattered gamma quanta varies from event to event. It can be however estimated
numerically based on simulations descried in this article.
The most relevant for the reconstruction of the hit-position along the scintillator strip is a difference 
(\(\Delta x\)) of \(x\)-coordinates between primary and secondary interaction points. 
\(x\)-direction is the one along the largest dimension of a scintillator strip. 
\begin{figure}[!h]
\centering
\includegraphics[width=0.6\textwidth]{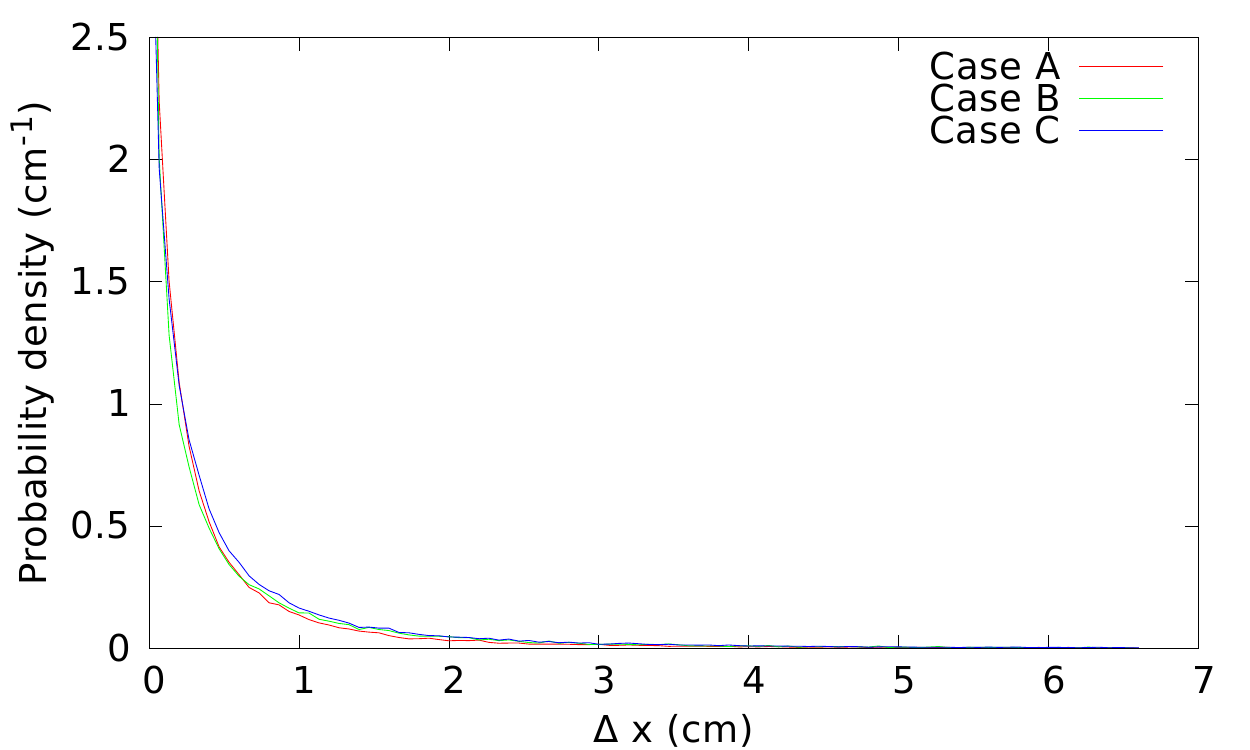}
\caption{Probability density functions of 
\(|\Delta x|\) defined as a distance along the scintillator between primary and secondary interaction points.
\label{abs}
}
\end{figure}
The greater 
\(|\Delta x|\) 
is, the more blurred is the signal from photomultiplier. 
A Full Width at Half Maximum of this distribution is rather small 
(FWHM(\(|\Delta x|\)) $\approx$ 0.3~cm). However since the distribution has a long tail
it is 
useful to tabularize its quantiles (see Tab.~\ref{tab:quantiles2}).
\begin{table}[!h]
\centering
\begin{tabular}{|c|c|c|c|}
\hline
Quantile & Case A & Case B& Case C \\\hline\hline
0.25 & \(0.042\unit{cm}\) & \(0.035\unit{cm}\) & \(0.060\unit{cm}\)\\
0.50 & \(0.16 \unit{cm}\) & \(0.16\unit{cm}\) & \(0.21\unit{cm}\)\\
0.75 & \(0.46 \unit{cm}\) & \(0.54\unit{cm}\) & \(0.61\unit{cm}\)\\
0.90 & \(1.1 \unit{cm}\) & \(1.3\unit{cm}\) & \(1.4\unit{cm}\)\\
0.95 & \(1.7 \unit{cm}\)  & \(2.1\unit{cm}\) & \(2.2\unit{cm}\)\\
0.99 & \(3.5 \unit{cm}\)  & \(3.8\unit{cm}\) & \(3.9\unit{cm}\)\\\hline
\end{tabular}
\caption{Quantiles of \(|\Delta x|\) distribution shown in Fig.~\ref{abs}.
\label{tab:quantiles2}}
\end{table}
The table indicates that more than 90$\%$ of secondary scattering occurs 
for \(|\Delta x|\) smaller than 1.4~cm. 
However, for the distortion of the signals not only a distance between interaction points is important,
but also the energy deposited by the secondary ionisation. 

\subsection{Energy deposition versus spatial separation of interaction points}
\begin{figure}[H]
\centering
\includegraphics[width=.6\textwidth]{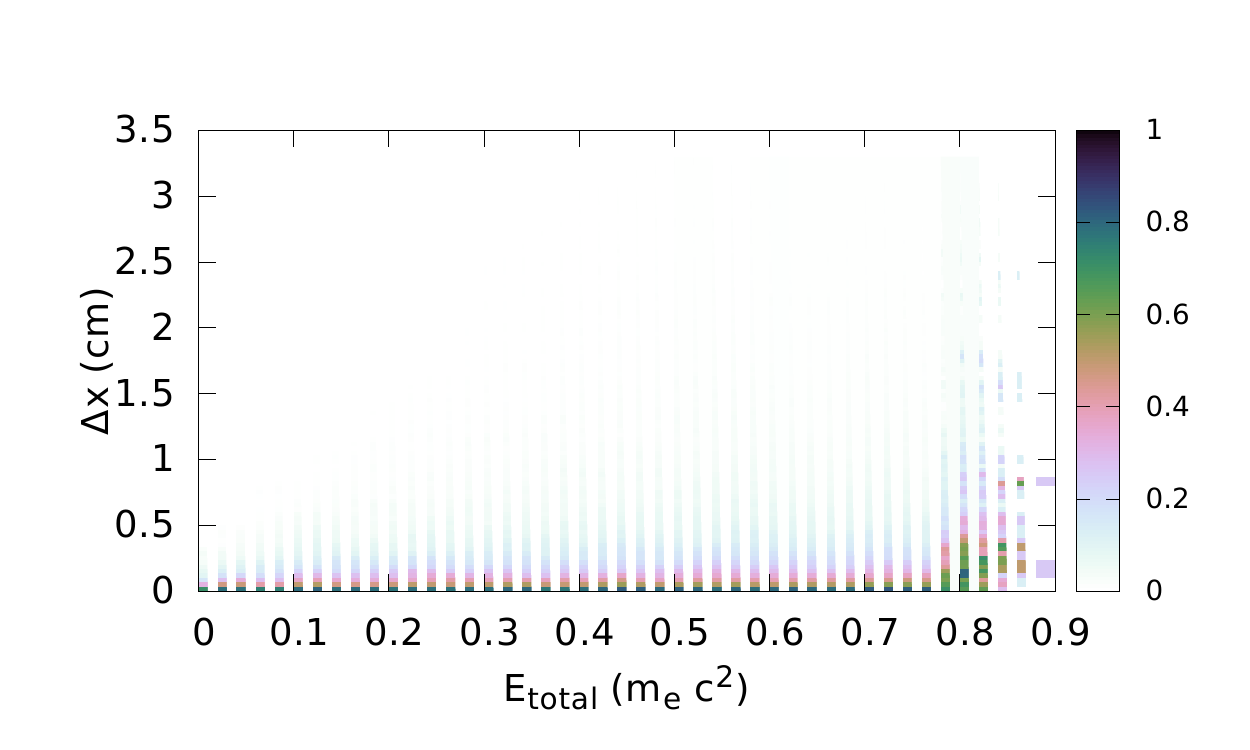}
\caption{Case A: Histograms of \(|\Delta x|\) as a function of total deposited energy. 
Note that this figure shows a set of one dimensional histograms normalized to unity in maximum.  
A more detailed description can be found in the text.}
\label{fig:edxa}
\end{figure}

\begin{figure}[H]
\centering
\includegraphics[width=.6\textwidth]{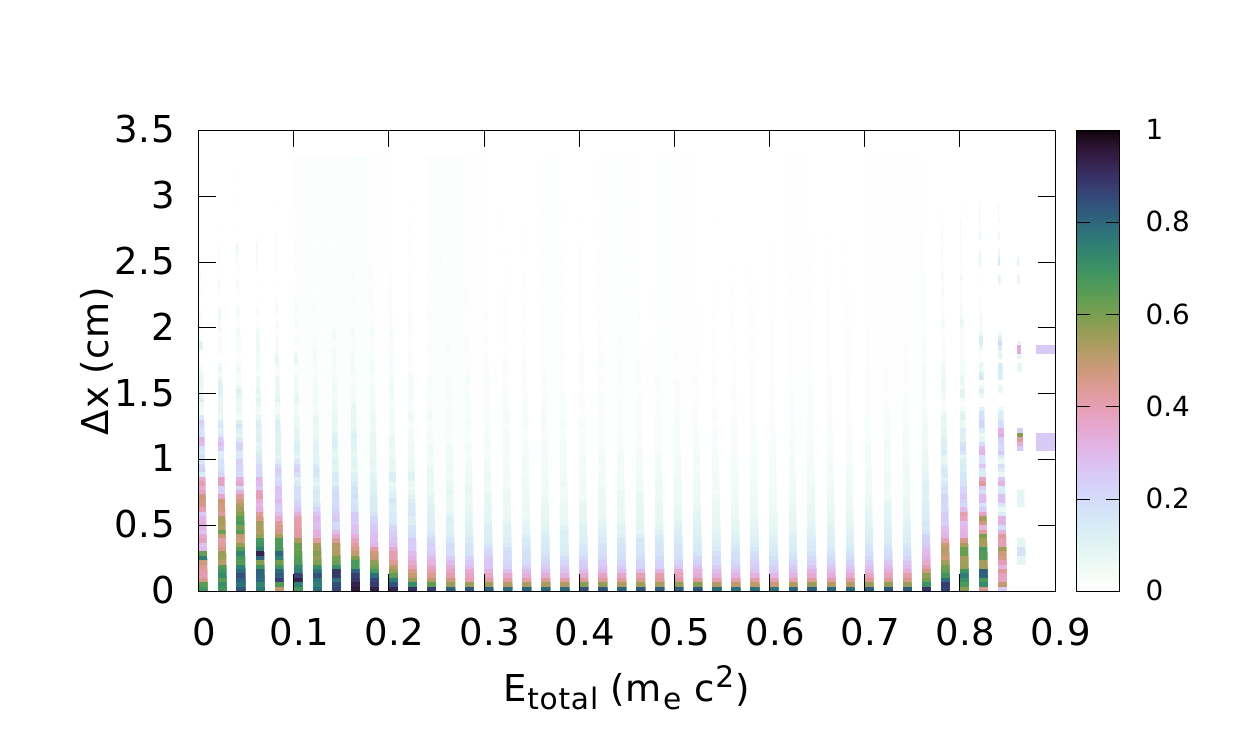}
\caption{Case C: Histograms of \(|\Delta x|\) as a function of total deposited energy. Note that this figure shows a set of one dimensional histograms normalized to unity in maximum.  A more detailed description can be found in the text.}
\label{fig:edxc}
\end{figure}

\begin{figure}[H]
\centering
\includegraphics[width=.6\textwidth]{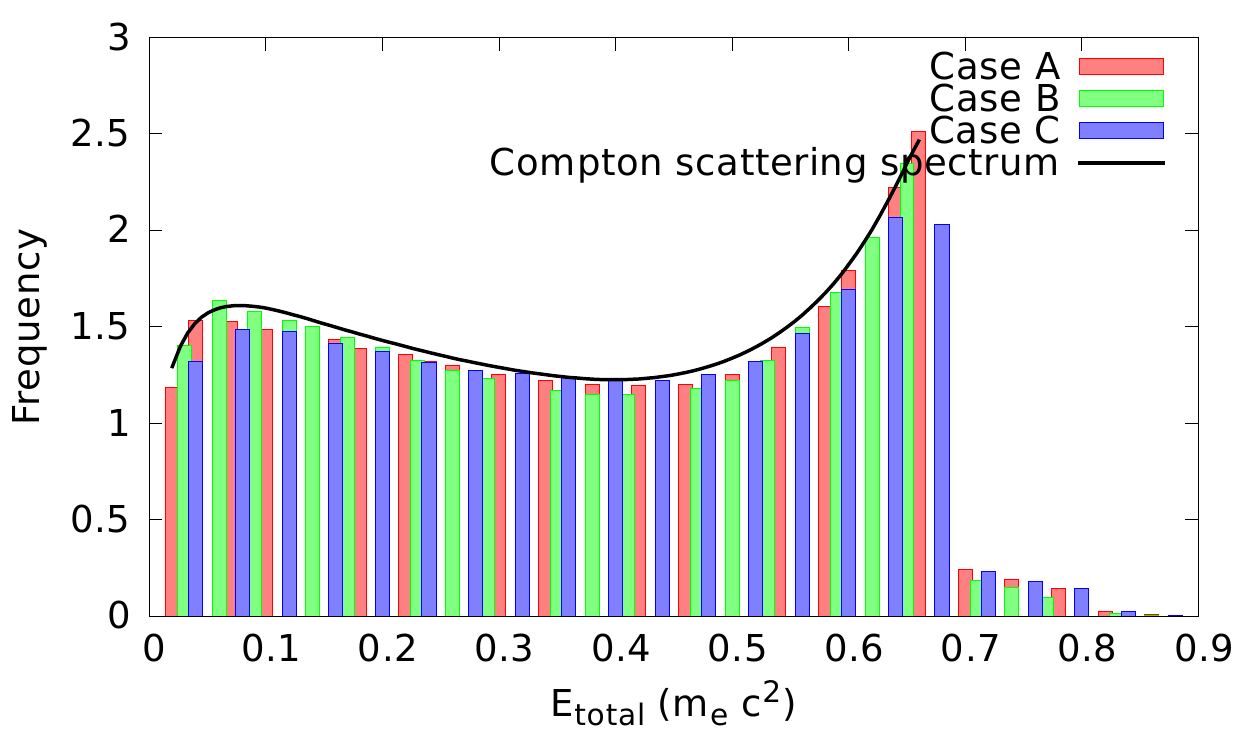}
\caption{Simulated spectra of total energy deposited in the scintillator strip 
by annihilation gamma quanta for case A, B and C. 
Superimposed solid line shows spectrum expected in case of a single scattering.}
\label{tote}
\end{figure}

Correlations between deposited energy and spatial separation of interaction points
is shown in figures~\ref{fig:edxa}--\ref{fig:edxc}.
These figures are however not 2--D histograms (i.e. heat maps), 
but sets of dense packed 1--D histograms, 
in which each line (sub--histogram) is separately normalized in that way, that maximum 
value in each sub--histogram is equal to \(1\). 
The figures are made this way, because one may approximate total energy stored 
in scintillator, and could be interested in most probable \(|\Delta x|\) in this event. 
It is important to mention that in this and following paragraph 
only events including at least two scatterings were taken into account.
\begin{figure}[H]
\centering
\includegraphics[width=.6\textwidth]{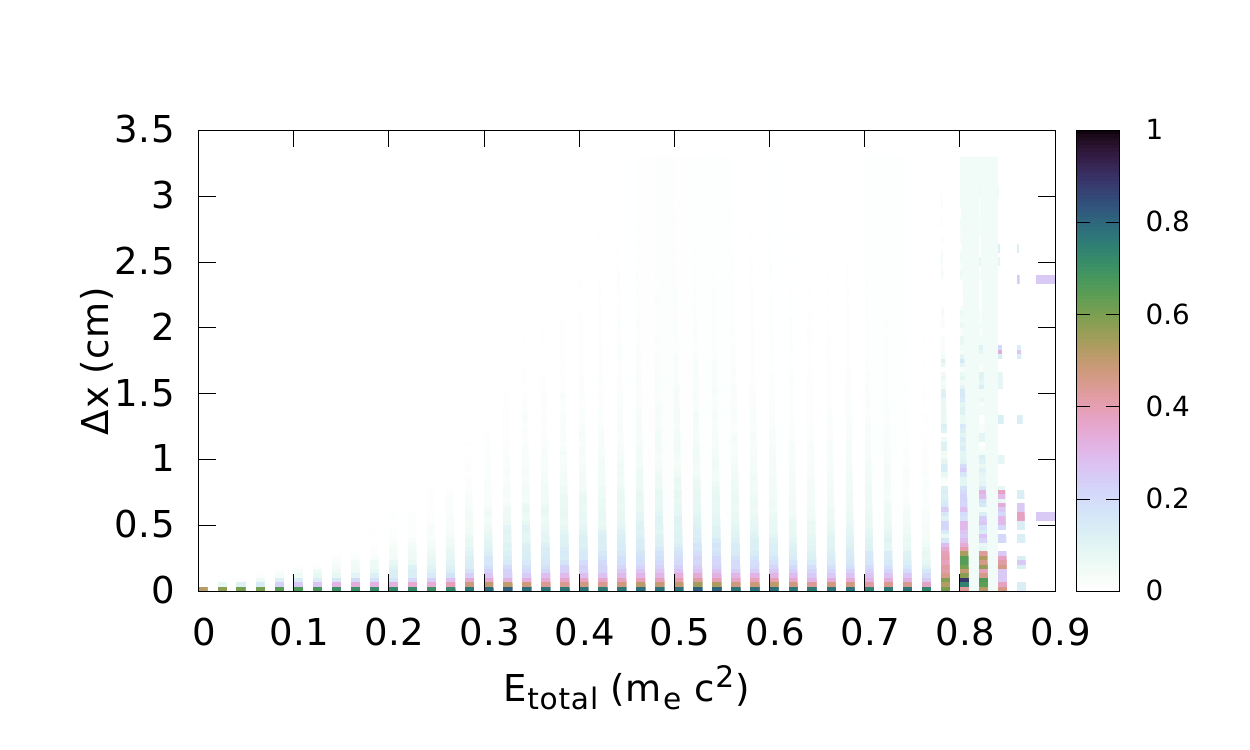}
\caption{Case B: Histograms of \(|\Delta x|\) as a function of total deposited energy. 
Note that this figure shows a set of one dimensional histograms normalized to unity in maximum.  
A more detailed description can be found in the text.}
\end{figure}
Results presented in these figures indicate that in order to decrese the blurring of the signals due to the secondary scattering
one can select from the full spectrum only these events for which deposited energy is larger than e.g.~0.2~\(m_e\) and smaller than 0.7~\(m_e\).   
For each of the studied cases the broadest 
\(|\Delta x|\) distributions are observed for the energy deposition larger 
than the maximum energy deposition in the primary scattering (0.67~\(m_e\)). 
In general the largest spread of 
\(|\Delta x|\) 
is observed as expected for the case C, 
for which the primary gamma quantum can travel the longest distance along the scintillator. 
In this case a scattering under small angles resulting in a small energy depositions leads to the scatterd gamma quantum 
which, due to the relatively large energy, can travel on the average large distance before the second scattering, which 
may again occur at most probable under a small angle with a small energy deposition. 
Therefore in this case broad 
\(|\Delta x|\) 
distributions are observed also at small values of total deposited energy.
The full energy spectra for all cases are presented in Fig.~\ref{tote}.

\subsection{Distribution of time differences between subsequent interactions}
\begin{figure}[H]
\centering
\includegraphics[width=.6\textwidth]{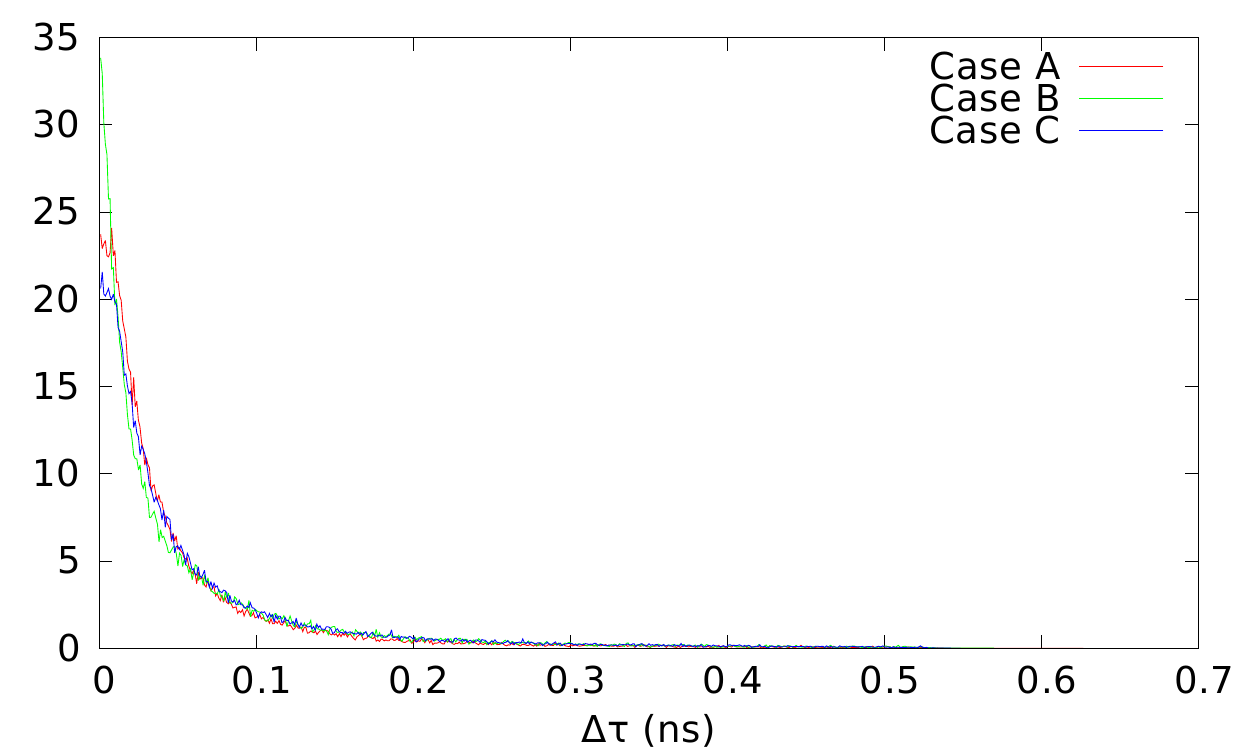}
\caption{\(\Delta \tau\) histogram (normalized to probability density function) for case A, B and C.}
\label{fig:dtau}
\end{figure}

Spatial distance of first and subsequent interactions, along with their time difference leads 
to blurring of signal coming to photomultipliers.
The light signals are read out by photomulitpliers on both edges of the strip,
therefore for one of them the signal distortion by the secondary scattering will 
be much smaller than for the other one.
This is because, between primary and secondary scattering the gamma quantum propagates towards one of the photomulitpliers
together with the primary light signal and therefore the signal distortion in this photomultiplier
will be smaller than in the other edge where the delay between primary and secondary light pulses 
is equal to:
\begin{equation}
\Delta \tau= \frac {\Delta x} {c'}  + \Delta t,
\end{equation}
where \(\Delta x\) denotes difference of x--component of interactions' positions, 
\(\Delta t\) denotes their time difference and \(c'\) stands for the effective velocity 
of light signal propagation trough the scintillator. In this calculations \(c'\)~=~14~cm/ns
was used.  
Histogram of \(\Delta \tau\) (normalized to PDF) is shown in Fig. \ref{fig:dtau}. 
Fig.~\ref{fig:dtau} shows that the primary and secondary light signals overlap with a delay of about 40~ps (FWHM of the distribution).

\section{Conclusions and discussion}
Simulations of scattering of annihilation gamma quanta 
in a strip of polymer scintillator have been conducted. Such strips constitute
basic detection modules of the newly proposed PET detector being developed by the J-PET Collaboration.
An algorithm simulating chain of Compton scatterings was elaborated and series of simulations have been conducted
for the scintillator strip with the cross section of  5~mm x 20~mm.
Simulations were simplified due to the observation that for the energy range of interest the Compton scattering 
is dominant and photoelectric and Rayleigh effects may be neglected. 
As a result:  (i) distributions of multiplicity of interactions,  
(ii) spatial distribution of interactions points as a function of the deposited energy,
and (iii) spectra of time differences between subsequent interactions
have been determined.
Obtained results indicate that secondary interactions occur only in the case of about 8$\%$ of events
and out of them only 25$\%$ take place in the distance larger than 0.5~cm.  It was also established that 
light signals produced at primary and secondary interactions overlap with the delay
which is spread by about 40~ps (FWHM). Moreover, analysis of histograms of the distance between subsequent interactions points
as a function of total deposited energy revealed that the blurring of signals due to the secondary interactions
may be decreased by 
selecting from the full spectrum only these events for which deposited energy is e.g. larger than \(0.2 m_e\) and smaller than \(0.7 m_e\).   

\section{Acknowledgements}
We acknowledge technical and administrative support by M. Adamczyk, 
T. Gucwa-Ry{\'s}, A. Heczko, M. Kajetanowicz,
G. Konopka-Cupia{\l}, J. Majewski, W. Migda{\l}, A. Misiak, 
and the financial support by the Polish National Center for Development 
and Research through grant INNOTECH-K1/IN1/64/159174/NCBR/12, 
the Foundation for Polish Science through MPD programme 
and the EU and MSHE Grant No. POIG.02.03.00-161 00-013/09.


\end{document}